\documentclass[sigconf]{acmart}
\usepackage{booktabs} 
\usepackage{multirow}

\usepackage{flushend}
\begin{document}
\title{On the Reproduction of Real Wireless Channel Occupancy in ns-3}

\author{Renato Cruz}
\affiliation{%
  \institution{INESC TEC and Faculdade de Engenharia, Universidade do Porto}
  \city{Porto}
  \country{Portugal}
}
\email{renato.m.cruz@inesctec.pt}

\author{Helder Fontes}
\orcid{0000-0002-7672-8335}
\affiliation{%
  \institution{INESC TEC and Faculdade de Engenharia, Universidade do Porto}
  \city{Porto}
  \country{Portugal}
}
\email{helder.m.fontes@inesctec.pt}

\author{Jos\'e Ruela}
\affiliation{%
  \institution{INESC TEC and Faculdade de Engenharia, Universidade do Porto}
  \city{Porto} 
  \country{Portugal}
}
\email{jose.ruela@inesctec.pt}

\author{Manuel Ricardo}
\orcid{0000-0003-1969-958X} 
\affiliation{%
  \institution{INESC TEC and Faculdade de Engenharia, Universidade do Porto}
  \city{Porto} 
  \country{Portugal}
}
\email{manuel.ricardo@inesctec.pt}

\author{Rui Campos}
\orcid{0000-0001-9419-6670} 
\affiliation{%
  \institution{INESC TEC and Faculdade de Engenharia, Universidade do Porto}
  \city{Porto} 
  \country{Portugal}
}
\email{rui.l.campos@inesctec.pt}

\renewcommand{\shortauthors}{R. Cruz, H. Fontes, J. Ruela, M. Ricardo and R. Campos}

\begin{abstract}

In wireless networking R\&D we typically depend on simulation and experimentation to evaluate and validate new networking solutions. While simulations allow full control over the scenario conditions, real-world experiments are influenced by external random phenomena and may produce hardly repeatable and reproducible results, impacting the validation of the solution under evaluation. Previously, we have proposed the Trace-based Simulation (TS) approach to address the problem. TS uses traces of radio link quality and position of nodes to accurately reproduce past experiments in ns-3. Yet, in its current version, the TS approach is not compatible with scenarios where the radio spectrum is shared with concurrent networks, as it does not reproduce their channel occupancy. 

In this paper, we introduce the \textit{InterferencePropagationLossModel} and a modified \textit{MacLow} to allow reproducing the channel occupancy experienced in past experiments. To validate the proposed models, the network throughput was measured in different experiments performed in the w-iLab.t testbed, controlling the channel occupancy introduced by concurrent networks. The experimental results were then compared with the network throughput achieved using the improved TS approach, the legacy TS approach, and pure simulation, validating the new proposed models and confirming their relevance to reproduce experiments previously executed in real environments.

\end{abstract}

\keywords{ns-3, Mobile Network Simulation, Trace Based Simulations, Reproducibility of Experimental Conditions, Perpetuation of Real-World Mobile Testbeds, Offline Experimentation, Channel Occupancy}

\maketitle

\section{Introduction}

Wireless networking R\&D depends on simulation and experimentation to evaluate and validate new networking solutions. While simulations allow full control over the scenario conditions (e.g., number of nodes, propagation loss, mobility, duration of experiment, obstacles), real-world experiments are influenced by external random phenomena (e.g., noise, interference, multipath, channel occupancy) and may produce hardly repeatable and reproducible results. The lack of repeatability and reproducibility directly impacts the validation of the solution under evaluation. 

Motivated by our hands-on experience with testbeds operating in emerging scenarios such as aerial \cite{sunny_project} and maritime \cite{bluecom+}, we have been developing  the Trace-based Simulation (TS) approach to enable repeatable and reproducible wireless networking experimentation using ns-3 \cite{Fontes2017}\cite{Fontes2018}. The TS approach introduces new mechanisms to 1) capture the execution conditions of an experiment and 2) enable its repetition and reproduction using ns-3 by relying on its good simulation capabilities from the MAC to the application layer. The results so far, including a large scale evaluation using the Fed4FIRE+ \cite{fed4fire} community wireless testbeds \cite{Lamela2019}, show that the TS approach is valid and achieves accurate reproduction of past experiments, even if the testbed becomes unavailable. The TS approach offers a special advantage over Pure Simulation (PS) approach in mobile scenarios, where the propagation loss characteristics are constantly changing and it is not possible to fine-tune a traditional \textit{PropagationLossModel} to be accurate throughout the movement of the nodes. Nevertheless, the TS approach also has limitations \cite{Fontes2019} such as the inability to reproduce the channel occupancy caused by concurrent radio transmissions from nodes that were in radio range but did not belong to the experiment we are willing to reproduce. This is a especially relevant limitation considering that in real scenarios it is very common to find multiple overlapping Wi-Fi networks.

The original contributions of this work are two-fold: 1) the improved version of the TS approach, now using a new ns-3 \textit{InterferencePropagationLossModel} and an updated \textit{MacLow} to reproduce the Channel Occupancy caused by other nodes not belonging to the experiment; 2) further validation of the TS approach, now focusing on non-controlled environments where reproducing the channel occupancy in ns-3 (on top of the radio link quality and mobility of nodes) is essential to maintain an accurate reproduction of past experiments.

The paper is structured as follows. In Section 2 we present the Related Work that directly influenced the contributions presented in this paper. In Section 3 we present the new \textit{InterferencePropagationLossModel} and the updated \textit{MacLow}, including their implementation details, in order to reproduce the channel occupancy of past experiments. In Section 4 we evaluate the new version of the TS approach, based on the models presented in Section 3, through experiments with controlled channel occupancy executed in the w-iLab.t testbed. Finally, in Section 5 we draw the conclusions and refer the future work.

\section{Related Work}

The work presented in this paper is directly related to previous publications addressing the repeatability and reproducibility of past experiments using the TS approach, which is based on trace-based ns-3 simulation models. 

In \cite{Fontes2017} the concept of TS approach was presented for the first time to reproduce a 2000 seconds long wireless networking experiment between an Unmanned Aerial Vehicle (UAV) and a Base Station (BS). The UAV flew over the sea up to 7 km away from the BS and at speeds of up to 400 km/h. The high cost of operation, weather constraints, and all the complex logistics involved were seriously impairing the communications solution evaluation. On top of that, because of the complex characteristics of the scenario, the pure simulation models of ns-3 were not able to accurately simulate those conditions. By recording traces of link quality (Signal-to-Noise ratio) and the position of nodes, and using the proposed \textit{TraceBasedPropagationLossModels}, the authors showed that it was possible to accurately repeat and reproduce the experiment in ns-3 without access to the testbed. The TS approach helped not only to improve the networking solution, but also other components that depended on communications to work such as the data publishing components on-board the UAV and the subscriber components connected to the BS node. By then, the TS approach supported point-to-point radio links, an SNR sampling resolution of 1 per second, and was evaluated using a fixed PHY-rate (6 Mbps) link.

In \cite{Fontes2018} the TS approach was evolved to support multiple-access scenarios. The improved \textit{TraceBasedPropagationLossModel} was tested in a laboratory testbed, showing that even in controlled environments the TS approach presented clear benefits over pure simulation when reproducing past experiments. This was mainly due to the non-negligible radio link asymmetry and clear RX sensitivity and TX power differences between nodes using identical Wi-Fi cards. In this paper it was also shown that using High SNR Sampling Rate (1 SNR sample per packet received instead of 1 SNR sample per second) was vital to reproduce the throughput more accurately when using auto-rate adaptation (Minstrel). Although the paper showed very promising preliminary evaluation results of the improved TS approach, a large scale evaluation was lacking.

To overcome this limitation, the large-scale evaluation of the TS approach using the community wireless testbeds made available by the Fed4FIRE+ federation was considered \cite{fed4fire} as part of the SIMBED F4Fp-OC3 European research project. Hundreds of Wi-Fi experiments were performed in point-to-point and multiple-access scenarios, as well as considering static and mobile nodes. Those experiments were then reproduced in ns-3 using the TS approach. The results \cite{Lamela2019} showed the clear value of the TS approach for accurately reproducing the past experimental conditions. However, some limitations were identified and a plan was devised in \cite{Fontes2019} to overcome them towards a better simulation-experimentation synergy using ns-3.

This work focuses on overcoming the TS approach limitation regarding the reproduction of the channel occupancy \cite{Fontes2019} experienced in the past real experiments, building on top of the proven strengths of the TS approach. Supporting the reproduction of channel occupancy has become increasingly important in Wi-Fi networks, due to their overlapping deployments, namely in dense urban scenarios. In these scenarios, multiple concurrent networks share overlapping spectrum, therefore interfering between themselves and significantly lowering the expected throughput. Even if the ns-3 pure simulation models are perfectly fine-tuned to reproduce the real RF environment conditions, the channel occupancy alone can significantly affect the performance evaluation, resulting in incomparable results between simulation and experimentation. To the best of our knowledge, the work presented herein is the first attempt to reproduce the channel occupancy of concurrent networks in simulation and represents a clear improvement over the previous version of the TS approach.

\section{Channel Occupancy Model} 
\subsection{Background}
Following the work in \cite{Fontes2017} and \cite{Fontes2018}, the proposed model builds on top of the same trace-based simulation principals, now focusing on supporting scenarios with shared radio spectrum with concurrent networks. 
Channel occupancy is introduced by nodes from other networks occupying the shared medium while our network is operational and also attempting to transmit. This concurrent spectrum usage have 2 main impacts in our transmission: 1) at the sender, where its transmission will be hindered when attempting to access the medium; 2) at the receiver, where hidden nodes may cause interference in the transmitted packet, causing it to be lost. For the sake of completeness, we now refer to the goal of the channel occupancy model and the most relevant aspects taken into account in its design.

\begin{itemize}

\item \textbf{Goal.} The goal of this work is to measure the real world channel occupancy caused by non-controlled networks, which influence the performance of our network, and reproduce this behaviour in ns-3 trace-based simulations. As the TS approach already accurately simulates and reproduces the performance of our network without external interference, this work only focuses on reproducing the impact of outside networks in our own network. Because outside networks are unknown and not controllable, they cannot be simulated alongside our simulation scenario.

\item \textbf{Variables to be collected.} In order to reproduce real world channel occupancy, each node of the network needs to measure how long they sensed the medium busy by other networks, and store these measurements in a "busy" trace file. Real wireless cards from Qualcomm Atheros (ath9k/ath10k-based drivers) report three values: 1) the total time the wireless card has been active; 2) the time the wireless card spent transmitting; and 3) the time the wireless card sensed the medium busy (either by its network or by any concurrent networks).

\item \textbf{Reproducing channel occupancy at the Sender.} Channel occupancy at the sender is caused by nodes from other networks attempting to transmit concurrently with our network, reducing the availability of the shared medium. This interference affects the sender node when attempting to gain access to the medium at the Mac layer. To implement this behavior, we modified the "Mac Low" to reproduce a busy medium according to the sender's channel occupancy observed in the real experiment.

\item \textbf{Reproducing channel occupancy at the Receiver.} Radio interference results in channel occupancy at the receiver. Such interference is introduced by nodes that are hidden from the transmitter, causing packets to be lost. Due to these losses, which are not expected by the sender, because it sensed the shared medium as free, a new propagation loss model was developed, the \textit{InterferencePropagationLossModel}, which blocks packet reception according to the receiver's channel occupancy observed in the real experiment.

\item \textbf{Channel Occupancy Simulation Settings.} As in \cite{Fontes2018}, the following settings should be taken into account: 1) TX Power End, TX Power Start and TX Gain; 2) RX Gain; 3) WiFi Standard, WiFi Mac, Frequency, Channel BW and Remote Station Manager; 4) Propagation Delay; 5) Propagation Loss; 6) Error Rate Model; 7) Data Mode and Control Mode and 8) Mobility Model. Furthermore, to reproduce channel occupancy, the proposed propagation loss model should be used, as well as the modified Mac layer. Ns-3 uses a linked list of propagation loss models, so it is advised that the \textit{InterferencePropagationLossModel} is used at the end of this chain to effectively block packet reception, as expected.
\end{itemize}

\subsection{Channel Occupancy}
Channel Occupancy is defined here as the proportion of time a node sensed the medium occupied by networks other than the Wi-Fi network it is associated to (current network). To measure it, each node's wireless card reports 1) the time it spent active (\textit{active}), 2) the time it sensed the medium busy, either by the current network or other networks (\textit{busytotal}) and 3) the time it spent transmitting (\textit{$t_x$}). Assuming that the receiver and sender are in radio range, the sender's \textit{$t_x$} will be accounted in the receiver's \textit{busytotal}. Furthermore, assuming all the nodes of the current network are in radio range of each other, the total transmission time of the current network will be accounted in the \textit{busytotal} of all nodes, so the remaining \textit{busytotal} is considered to be the transmission time from other networks sensed at node $i$ ($busyother_i$), defined by Equation \ref{eq:eq1}, where $n$ represents the number of nodes in the current network.
\begin{equation}
    busyother_i = busytotal_i - \sum^n_{\substack{
                                    j=0\\
                                    i\neq j}}(t_{x_j}) 
\label{eq:eq1}
\end{equation}

Each node's Channel Occupancy (\textit{CO}) is calculated every 1000 ms using Equation \ref{eq:eq2}: 

\begin{equation}
    CO_i = \frac{busyother_i}{1000}
\label{eq:eq2}
\end{equation}

\subsection{Receiver's Channel Occupancy}

\begin{figure}[h]
\centering
\includegraphics[width=\linewidth]{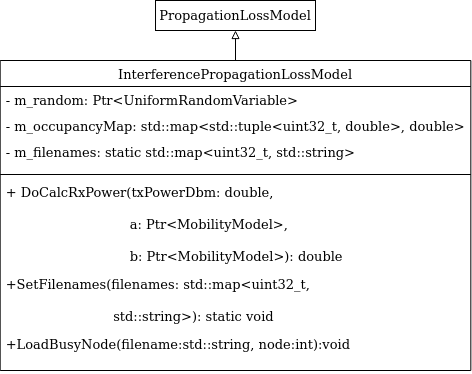}
\caption{Class diagram of the \textit{InterferencePropagationLossModel}.}
\label{fig:ClassDiagram}
\end{figure}

As previously stated, to reproduce the receiver's channel occupancy, a new propagation loss model was developed (\textit{InterferencePropagationLossModel}), whose class diagram is represented in Figure \ref{fig:ClassDiagram}. The \textit{InterferencePropagationLossModel} model maps every simulated node's channel occupancy calculated using Eq. \ref{eq:eq2} to simulation time. This is implemented with the following modifications to the \textit{PropagationLossModel} class:

\begin{itemize}
    \item \textbf{m\_filenames} - static map to store each node's "busy" trace file name, generated during the real experiment. It needs to be set during simulation setup using the method \textbf{SetFilenames}, where each file name is mapped to the node id. As this map is static, every instance of the propagation model uses it during its initialization to process the files of each node and map the channel occupancy samples to simulation time in \textbf{m\_occupancyMap}.
    
    \item \textbf{m\_occupancyMap} -  maps each <\textit{node id},\textit{time}> pair to a channel occupancy sample, initialized in the model's constructor with \textbf{LoadBusyNode}, using the \textit{node id} file name stored in \textbf{m\_filenames}.
    
    \item \textbf{m\_random} - uniform random variable used to decide if the receiver can receive the transmitted packet. In each \textbf{DoCalcRxPower} call, if the receiver has a free medium, then the transmission power (\textit{txPowerDbm}) is returned, so the transmission remains unaffected. Otherwise, with an occupied medium (interference/collision), a low reception power (\textit{rxPowerDbm}) is returned, so that the receiver is unable to properly decode the transmitted packet, blocking the transmission. This behaviour is described in Equation \ref{eq:eq3}.
    \begin{equation}
        rxPowerDbm = m\_random < CO_i ? -1000 : txPowerDbm   
        \label{eq:eq3}
    \end{equation}
    
\end{itemize}

\subsection{Sender's Channel Occupancy}
\begin{figure}[h]
\centering
\includegraphics[width=0.8\linewidth]{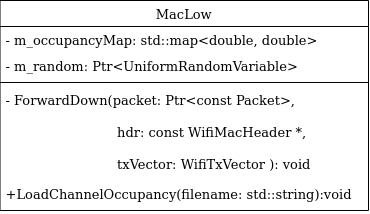}
\caption{Class diagram of the modified \textit{MacLow}.}
\label{fig:MacClassDiagram}
\end{figure}

Figure \ref{fig:MacClassDiagram} depicts the modifications made to the \textit{MacLow} layer of ns-3. A similar approach to the \textit{InterferencePropagationLossModel} was taken, where each channel occupancy sample is mapped to simulation time:

\begin{itemize}
    \item \textbf{m\_occupancyMap} - maps each channel occupancy sample with the corresponding simulation time. The only difference from the \textit{InterferencePropagationLossmodel} m\_occupancyMap, is that each node has its own \textit{MacLow} layer, so there is no need to store the node id to which each sample belongs to. This map must be initialized during simulation setup using \textbf{LoadChannelOccupancy}, passing the "busy" trace file name as parameter. 
    
    \item \textbf{m\_random} - the \textbf{ForwardDown} function was modified to access the \textbf{m\_occupancyMap} to decide if a packet should be forward to the physical layer or blocked, using this uniform random variable in a similar way as described in the \textit{InterferencePropagationLossModel}.
\end{itemize}

Although this model was implemented in ns-3.27, as this work is a continuation of the work presented in \cite{Fontes2018}, implemented over ns-3.27, the channel occupancy model described herein is also compatible with more recent versions of ns-3. The developed code can be found in \cite{ns-3_channel_occupancy_model}, with instructions on how to deploy and use the new model in ns-3.27.

\section{Evaluation of the Channel Occupancy Model}
To validate the new model and the implemented modifications in ns-3, two tests were performed: 1) simulation based on traces composed by synthetic predefined channel occupancy values, in order to assess whether the results obtained in ns-3 using the new model are the expected ones with respect to the enforced channel occupancy values (e.g., for a channel occupancy of 50 \% we expect near 50 \% of the maximum throughput); 2) using a Fed4FIRE+ testbed \cite{fed4fire} and considering a \textbf{controlled scenario} with two Wi-Fi communicating nodes (the experiment we want to reproduce in ns-3), and an outside Wi-Fi network causing controlled interference (the one that is accounted in the channel occupancy/"busy" trace files). In each simulation, the sender's and receiver's channel occupancy were first tested independently, one simulation using only the \textit{InterferencePropagationLossModel} and another using only the modified \textit{MacLow}. Then, both models were tested together.

\subsection{Synthetic Data Test}
To test each implementation separately, a test was devised comprised of two nodes (A and B), starting with unidirectional \textit{iperf} UDP flows, A->B and B->A, and then a bidirectional flow, A<->B, each flow with the following channel occupancy pattern:

\begin{itemize}
    \item Both nodes start with no channel occupancy;
    \item Every 5 s, the receiver's channel occupancy increases by 10\%, up to 50\% channel occupancy.
    \item After 5 s with 50\% channel occupancy, it is reset to 0\% channel occupancy for 5 s.
    \item Sender's channel occupancy starts increasing following the same pattern.
\end{itemize}

\begin{figure}[h]
\centering
\includegraphics[width=\linewidth]{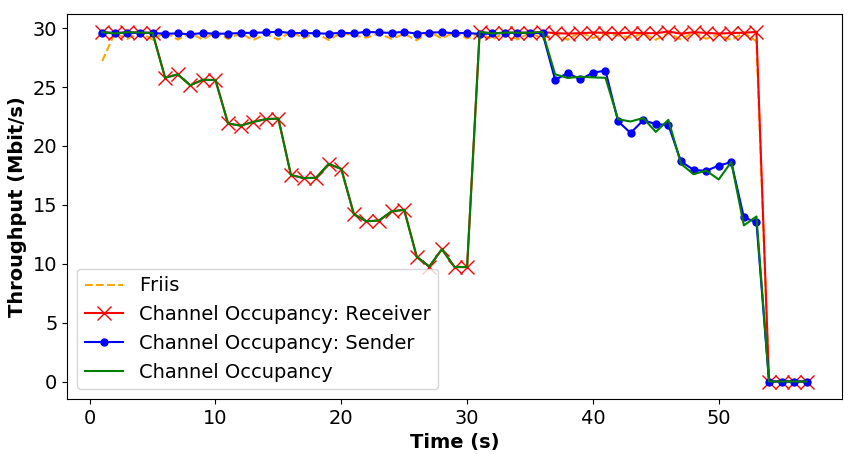}
\caption{UDP Throughput obtained with fixed communication parameters, testing the developed channel occupancy model on the receiver, sender and both with pre-defined channel occupancy values.}
\label{fig:ForgedData}
\end{figure}

Figure \ref{fig:ForgedData} presents the results of the Synthetic Data test, using UDP traffic, the IEEE 802.11a standard, a fixed rate of 54 Mbit/s, an SNR of 75 dB, a noise floor of -95 dBm, and a channel bandwidth of 20 MHz.

The results obtained using only the \textit{InterferencePropagationLossModel} (red line) show the impact of the increasing channel occupancy at the receiver, where the Throughput decreases until the 30 s mark in simulation time, following the channel occupancy pattern described before. After the 30 s mark, the channel occupancy at the receiver is set to 0\% until the end of the simulation, which means in this scenario (only the propagation loss model is being used), neither node (A or B) have channel occupancy imposed by other concurrent nodes, so the Throughput is the maximum possible.

Likewise, when testing the modified \textit{MacLow} layer (blue line), the Throughput also follows the channel occupancy pattern described for the sender. It starts with no channel occupancy (maximum Throughput) during the first 35 s of simulation. Then, the channel occupancy at the sender increases gradually, resulting in reduced Throughput until the end of the simulation.

Finally, results from both models tested together (green line), reproducing the channel occupancy at the sender (modified \textit{MacLow}) and receiver (\textit{InterferencePropagationLossModel}), show that for the first 30 s of simulation, where the receiver's channel occupancy increases gradually, the Throughput decreases accordingly. After this, for 5 s, neither node has channel occupancy, so the Throughput is the maximum possible. For the rest of the simulation, the Throughput decreases gradually with the increase of the sender's channel occupancy.

For comparison, the results obtained using \textit{Friis} propagation loss model (orange line) are included, which is equivalent to having no channel occupancy, so the Throughput is the maximum possible throughout the simulation.

With this test it was possible to confirm that the new channel occupancy model is able to influence the resulting Throughput according with the channel occupancy experienced by the nodes.

\subsection{Controlled Interference Experiment}
To further evaluate and validate the proposed channel occupancy model, a similar test was performed using Fed4Fire+ w-iLab.1 testbed \cite{fed4fire}, in order to verify the real impact that the channel occupancy has on a real communication and confirm that the same behaviour is reproduced in ns-3. The test was comprised of two nodes communicating within a private Wi-Fi network (i.e., the experiment we want to reproduce), using \textit{iperf} generating UDP traffic above link capacity to occupy the available bandwidth, while two outside nodes (i.e. the interfering nodes) communicating with each other, also using \textit{iperf} to generate channel occupancy within range. The traffic generated by the outside nodes introduces channel occupancy on both nodes operating in the private Wi-Fi network as they are all in range of each other (no hidden nodes), and follows a similar pattern as the previous test with synthetic data:

\begin{itemize}
    \item Both nodes in our network start with 0\% channel occupancy (no \textit{iperf} traffic from outside nodes).
    \item Channel occupancy on both nodes in our network increases 10\% every 5 s up to 50\% (\textit{iperf} traffic from outside network increases every 5 s). Because the maximum throughput in this radio link is ~30 Mbit/s, each 10\% increment of channel occupancy generated by the outside/interfering nodes corresponds to an increment of 3 Mbit/s in the UDP flow generated between the outside nodes.  
\end{itemize}

The current test was divided in three different experiments, accordingly to the direction of the UDP flow: 1) from node A to node B; 2) from node B to node A; and 3) bidirectional between node A and node B.

\begin{figure}[h]
\centering
\includegraphics[width=\linewidth]{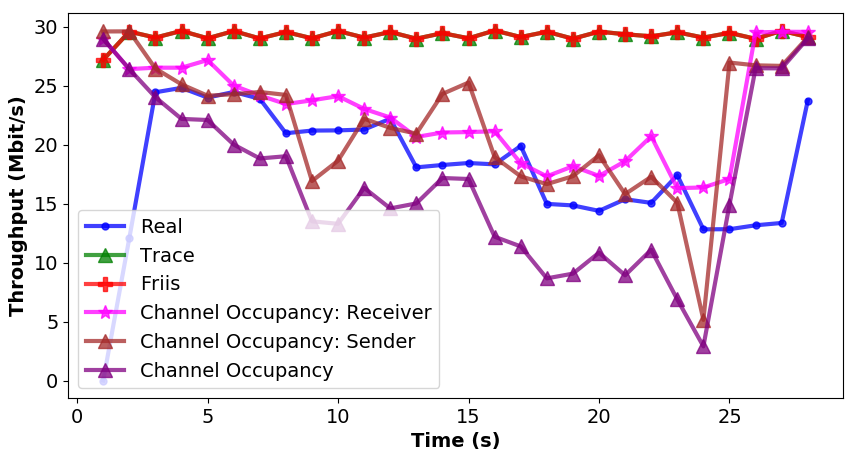}
\caption{UDP Throughput obtained with controlled interference from outside networks.}
\label{fig:testscenario}
\end{figure}

Figure \ref{fig:testscenario} shows the results from the first experiments (node A --> node B) of the test scenario with controlled interference as described, using the IEEE 802.11a standard, UDP traffic, channel bandwidth of 20 MHz, and both networks communicating concurrently in channel 36. As previously stated, the results for this experiment are slightly different from the synthetic data test as the channel occupancy is introduced at both the receiver and sender at the same time due to all the nodes being in radio range of each other.

Throughput from the real experiment (blue line) show the impact that the interfering traffic has on the current network. It decreases gradually as channel occupancy increases during the experiment. Furthermore, the same behaviour is observed in the simulation data using channel occupancy of the receiver with the propagation loss model (pink line), the sender with the modified Mac layer (brown line) and using both models (purple line). It can be seen that the proposed model leads to lower Throughput than the Throughput obtained in the real experiment. Results from the \textit{InterferencePropagationLossModel} (pink line) and the modified \textit{MacLow} (brown line) are similar as both nodes have the same channel occupancy (no hidden nodes). For comparison, results obtained using \textit{Friis} propagation loss model (red line) and the \textit{TraceBasedPropagationLossModel} (green line) \cite{Fontes2018}\cite{ns-3_trace_based_propapation_model2018} are included and, as shown, are similar as neither models take into account the channel occupancy, i.e., they simulate a "\textit{perfect}" communication. 

\begin{table*}[]%
\caption{Average UDP Throughput obtained in the controlled interference scenario.}
\label{tab:goodput}
\begin{tabular}{lllll|l|l|}
\cline{3-7}
                            & \multicolumn{1}{l|}{}                              & \multicolumn{3}{l|}{\textbf{Average UDP Throughput (Mbit/s)}}                            & \multicolumn{2}{c|}{\textbf{Relative Error}}         \\ \hline
\multicolumn{1}{|l|}{\textbf{Exp\#}} & \multicolumn{1}{l|}{\textbf{Flow}}                          & \multicolumn{1}{l|}{\textbf{Real Exp.}} & \multicolumn{1}{l|}{\textbf{ChOccup Exp.}} & \textbf{Friis Sim.} & \textbf{ChOccup Exp.}          & \textbf{Friis Sim.}          \\ \hline
\multicolumn{1}{|l|}{1}     & \multicolumn{1}{l|}{A -\textgreater B}             & \multicolumn{1}{l|}{17.92}     & \multicolumn{1}{l|}{16.67}        & 29.26      & 6.96 \%               & 63.28\%             \\ \hline
\multicolumn{1}{|l|}{2}     & \multicolumn{1}{l|}{B -\textgreater A}             & \multicolumn{1}{l|}{18.05}     & \multicolumn{1}{l|}{14.29}        & 29.25      & 20.98\%               & 65.50\%             \\ \hline
\multicolumn{1}{|l|}{3}     & \multicolumn{1}{l|}{A \textless{}-\textgreater{}B} & \multicolumn{1}{l|}{8.55}      & \multicolumn{1}{l|}{9.90}         & 13.48      & 7.53\%                & 27.57\%             \\ \hline
                            &                                                    &                                &                                   &            & \multicolumn{2}{l|}{\textbf{Average Relative Error}} \\ \cline{6-7} 
                            &                                                    &                                &                                   &            & 11.83\%               & 51.12\%             \\ \cline{6-7} 
\end{tabular}
\end{table*}

The average Throughput values measured in all the experiments, and their respective simulations counterparts, are presented in Table \ref{tab:goodput}. Note that the simulation using Friis produces equivalent results to the TS approach, as we are experimenting with a scenario with ideal link quality, and both models do not reproduce the channel occupancy. The first line (Exp \#1) shows the average throughput of the results already presented in Figure \ref{fig:testscenario}, while the others represent the other two experiments differing only on the traffic flow directions. Analysing the relative error of the average throughput of each simulation model when compared to the real experiment it is possible to conclude that average throughput is significantly closer to the real experiment, producing more accurate results. Considering the three experiments, on average, the throughput relative error was reduced from 51.12\% to 11.83\% which represents a significant improvement over pure simulation (Friis) and Trace-based simulation without channel occupancy reproduction.

\section{Conclusions and Future Work}

We developed an improved TS approach to reproduce channel occupancy in ns-3 by controlling both the sender and the receiver. At the sender, by manipulating its layer 2 access mechanism, simulating a busy medium; at the receiver, by implementing the \textit{InterferencePropagationLossModel}, a new propagation loss model to induce packet loss that replicates radio interference from nodes near the receiver that are hidden from the transmitter. 

Two tests were performed to evaluate the proposed model. Firstly, by feeding synthetic channel occupancy values to ns-3, which helped validating that the model was working as expected (e.g., with 20 \% channel occupancy, we would be able to get only about 80 \% of the expected throughput for the same link quality). Secondly, by using w-iLab.1, a Fed4FIRE+ community testbed, where experiments were performed with real nodes to assess the model correct operation using real "busy"/channel occupancy traces collected from real experiments. 

In w-iLab.1 controlled environment two pairs of nodes were used: one pair representing our network and the second pair representing the concurrent network, which we used to control the amount of channel occupancy induced. Using the TS approach with channel occupancy support we were able to accurately reproduce the throughput of real experiments, lowering the average relative error from 51 \% to 12 \%, when compared to the results obtained using pure simulation or the legacy TS approach, which do not account for channel occupancy when reproducing past experiments. 

As future work, we will continue to further improve the channel occupancy model in order to support IEEE 802.11ac and IEEE 802.11n. Furthermore, more experimentation is still needed in real world uncontrolled scenarios, with random channel occupancy introduced by uncontrolled networks. To do this we plan to use the Fed4Fire+ CityLab testbed, where the nodes are placed around the city of Antwerp, and are subject to an unknown number of outside networks, with unknown and uncontrolled channel occupancy. 

\begin{acks}
This work is financed by national funds through the FCT -- Foundation for Science and Technology, I.P., under the project: 
UID/EEA/\\50014/2019.

This work is also part of the SIMBED+ project, approved in Fed4FIRE+
Open Call 5.

\end{acks}
\bibliographystyle{ACM-Reference-Format}
\bibliography{refs}
\end{document}